\documentclass[aip,prl,reprint,floatfix]{revtex4-2}  
\usepackage{graphicx}  
\usepackage{dcolumn}   
\usepackage{bm}        
\usepackage{amssymb}   
 \usepackage{float}
\usepackage{amsfonts}
\usepackage{gensymb}
\usepackage{amsmath}
\usepackage{upgreek}
\usepackage{svg}
 \usepackage{textcomp}
 \usepackage{hyperref}

\hypersetup{
    colorlinks=true,
    linkcolor=blue,
    filecolor=cyan,      
    urlcolor=cyan,
}
 
\usepackage{listings}
\usepackage{color}

\definecolor{dkgreen}{rgb}{0,0.6,0}
\definecolor{gray}{rgb}{0.5,0.5,0.5}
\definecolor{mauve}{rgb}{0.58,0,0.82}

\usepackage{blindtext,tikz}
\usetikzlibrary{calc}
\nocite{*}
\begin{document}
\title{\textsc{SpectroLab}: An Open Source Matlab Based Toolbox for High Throughput Spectroscopy Analysis}
\author{Christopher Sims}
\affiliation{\textit{School of Electrical and Computer Engineering, Purdue University, West Lafayette, Indiana 47907, USA}}
\date{\today}

\begin{abstract}

We present an open source software package \textsc{SpectroLab} a Matlab based tool developed in 2018 for the analysis of spectroscopic data. In this package there are tools for derivative analysis, stacked energy contours , stacked plots for theory, 3D volumetric plots, and core level analysis, and derivative analysis. The package can currently be used for angle-resolved photoemission spectroscopy (ARPES) data, with the ability to also be used for other spectroscopic data in the future. We apply this program to the HfP$_2$, ZrSiS, and Hf$_2$Te$_2$P systems to demonstrate its robustness.

\end{abstract}
\maketitle
\section{Introduction}

In recent years, spectroscopic techniques such as angle resolved photoemission spectroscopy (ARPES)\cite{Damascelli2004,Lu2012}, scanning tunneling microscopy (STM) \cite{Binnig1982,Tersoff1983,Binnig1987}, neutron scattering \cite{Wilkinson1961,PRINCE2004,Strobl2017}, etc. have improved to the point where the scientific communities use these tools extensively. These techniques reveal features in momentum and energy space that can give deeper understanding of the interactions of exotic quasi-particles that have taken interest in recent years such as topological insulators\cite{Hasan2010,Xia2009a}, Weyl semimetals\cite{Xu2015,Lv2015,Huang2015}, nodal-line semimetals\cite{Burkov2011,Fang2015,Fang2016}, and High order Topological insulators (HOTI)\cite{Regmi2019,Schindler2018}. Furthermore, these techniques have been expanded to measure other dimensions such as spin, time, magnetism, etc. These techniques are always prone to noise which will decrease the ability to interpret the data correctly. Previously, researchers have employed proprietary programs in order to conduct the data analysis. These programs limit the ability for users to tune the way results are presented and create figures and graphs with a high degree of control. Here, we report a Matlab based program that is written in an easy to read way that allows for the analysis of ARPES data. Using Matlab's built in tools, we have design a program that gives users a high control in the way they present their data.

The data analysis of these materials is critical in order to understand complex interactions that are difficult to resolve by eye. By importing the data into Matlab, we provide a way for researchers to use the vast packages included in this program that are nontrivial to code in proprietary data analysis programs such as edge detection or neural networks. We present a suite of functions that allow for easy analysis and presentation of spectroscopic data in the Matlab suite called \textsc{spectrolab}. The suite needs minimal programming experience since all of the tools included can be done in one click.  As oppose to other data analysis techniques, our package is all inclusive and can be used with minimal need to modify the code in anyway. The only need for the program to work is an \emph{.ibw} file that is formatted correctly. We present ARPES data in from the HfP$_2$ \cite{Sims2019}, ZrSiS \cite{Neupane2016,Hosen2017}, and Hf$_2$Te$_2$P \cite{Hosen2018} systems in order to show that this program is applicable to different types of ARPES data. This paper includes a detailed description of the theory behind how the main functions work in section 2. In section 3, we outline the functionalities of each of the functions in the initial release. In section 4, we go over how to install and use \textsc{SpectroLab}. Finally in section 5, we go over a few examples of using the software package in real systems.

\section{Methods}
As the data is imported into \textsc{SpectroLab} it is automatically interpreted into $k$-space from the raw data using the following equations\cite{Comin2015}:
\begin{equation}
k_x = \frac{1}{\hbar}\sqrt{2mE_{kin}}sin\vartheta cos\varphi
\end{equation}
\begin{equation}
k_y = \frac{1}{\hbar}\sqrt{2mE_{kin}}sin\vartheta sin\varphi
\end{equation}
In the program these are interpreted by the following approximation
\begin{equation}
k_x = 0.5112*\sqrt{(E-4.5)}*(\pi*X*0.046/180)
\end{equation}
\begin{equation}
k_y = 0.5112*\sqrt{(E-4.5)}*(\pi*Z/180)
\end{equation}
In addition to this all supporting variables are automatically calculated within the setvariables() function.
\begin{lstlisting}
KY_Angle_Max = 0.5112*sqrt(E -4.5)*(pi*Zmax/180);
KY_Angle_Min = 0.5112*sqrt(E -4.5)*(pi*Zmin/180);
KX_Angle_Max = 0.5112*sqrt(E-4.5)*(pi*Xmax*0.046/180) -1.68;
KX_Angle_Min = 0.5112*sqrt(E-4.5)*(pi*Xmin*0.046/180) - 1.68;
\end{lstlisting}
\subsection{Sobel filter (2nd Derivative)}
The common edge detection algorithms are the Sobel, Canny, Prewitt, Roberts, and fuzzy logic methods\cite{Senthilkumaran2009,Ma2010,Gao2010,Seif2010,Mintz1994,Kanopoulos1988,Chen2012,Canny1986,Huertas1986,G.T.Shrivakshan}. In this paper, we use a modified version of the Sobel filter in order to conduct second derivative calculations. While it is possible to use a convolution neural network (CNN) for feature extraction\cite{Bishop1996,Peng2020}, these networks require a large amount of computational resources and are only applicable to the data set they are trained on. Generative adversarial neural networks (GANs)\cite{Goodfellow2014} are a much better alternative to create a neural network feature extraction since it can work on an infinite amount of data sets once it is properly trained. The downside is that a large amount of data sets ($>$1000) are required in order to properly train the neural network. Therefore, we forgo the inclusion of a neural network in our program. The Sobel filter is a 3$\times$3 matrix that calculates the second derivative locally in an image file (where A is the image). There are two versions, the horizontal and the vertical which can be used, in the distributed program. The horizontal $F_x$ is convoluted with the image to form the second derivative of the image.
\begin{equation}
F_x  = 
  \begin{bmatrix}
    -1 & 0 & +1 \\
    -2 & 0 & +2\\
      -1 & 0 & -1
   \end{bmatrix} * A , 
   F_y =
     \begin{bmatrix}
    -1 & -2 & -1 \\
    0 & 0 & 0\\
      +1 & +2 & +1
   \end{bmatrix} * A
\end{equation}
In order to detect edges after the Sobel filter is applied, one would find the square of the absolute value of the resultant $|F_x|^2$. By not conducting this last step we can resolve weak signals of ARPES. In the case of sharp data, the edge detection feature can be turned on with an additional line of code which we have included. All edge detection algorithms can be accesses via Matlab's image toolbox.
\begin{lstlisting}
Data2D = squeeze(Mapping(:,EnergyLevel,:))';
h_Sobel = fspecial('sobel');
f_y = imfilter(double(Data2D), h_Sobel,'replicate');
imagesc(f_y)
\end{lstlisting}
\subsection{Constant Energy Contours}
An energy contour is a slice of a 3D band mapping in ARPES ($k_x$,energy,$k_x$) where at a constant energy, the 2D $k_x$,$k_y$ surface is revealed. For constant energy contours we create a function that automatically interprets the energy levels and recreates them in an energy contour. For labeling we included a rounding function that rounds to the nearest significant figure. However, the program can interpret the exact value for publication.
\begin{lstlisting}
Energy_Limit = 200; %specified by user
Energy_Cut = 1:int16(Energy_Limit/6):Energy_Limit;
Energy_Contour = ypos - Energy_Cut % check limits;
hold on
subaxis(2,6,1,'sv',vertical_space,'sh',0.01); 
f = squeeze(B(:,ypos,:))';
imshow(imrotate(f,0),[], 'XData', [KX_Angle_Min+ + offsetx KX_Angle_Max+ offsetx]...
    , 'YData', [KY_Angle_Min + offsety KY_Angle_Max + offsety], 'colormap', setmap);
for ii = 2:6
subaxis(2,6,ii,'sv',0.01,'sh',0.01); 
f = squeeze(B(:,Energy_Contour(ii),:))';
imshow(imrotate(flipud(f),0),[], 'XData', [KX_Angle_Min KX_Angle_Max]...
    , 'YData', [KY_Angle_Min KY_Angle_Max], 'colormap', 'gray' );
    end
\end{lstlisting}
\subsection{Volumetric plot}
The volumetric plot function takes advantage of how data is formatted for ARPES. Typically data is encoded in an (X,Y,Z,C) format. Where X,Y,and Z are $k_x$, $k_y$, and energy, while C is the interpretation of the intensity at that location. By using an alpha map, we can make weak signals transparent and reveal the 3D band structure for visualization. This has the ability to be a very powerful tool for investigating anisotropy in bands.
\begin{lstlisting}
%VolumetricPlot()
% edit this to get a good mapping
alphamap([0 linspace(0.0005, 0.4, 255)]);
\end{lstlisting}
\subsection{Momentum cuts}
We have included several methods to aid the taking $k_x$ and $k_y$ momentum cuts from a 3D band mapping. We take these cuts by selecting a value of $k_x$ and $k_y$ and resolving the cut in a 2D image.
\begin{lstlisting}
f = squeeze(B(Kxlocation,:,:));
f = imresize(f,5);
imshow(f',[], 'XData', ...
    [KX_Angle_Min KX_Angle_Max], 'YData',...
    [abs(Ymin) - Fermi_Level, abs(Ymax)- Fermi_Level], 'colormap', setmap);
\end{lstlisting}
\subsection{Energy \& momentum distribution curves}
 Energy \& momentum distribution curves (EDC \& MDC) are used for the data analysis of spectroscopic data. In a set amount of intervals a line of data is taken out of the image and then plotted in order to enable researchers to resolve difficult to resolve images from ARPES or other spectroscopic data. We use the \emph{waterfall()} function in order to reconstruct the EDC and MDC plots seen in literature \cite{Hosen2018}. The plot also have the benefit of being three dimensional with the ability to add a color map to further define intensity differences.
\subsection{Stacked theory}
In order to import theory, we create a 2D surface in a 3D box using Matlab's \emph{meshgrid()} function to import an image of the theoretical plot. To insure that the images that we import are of the same size we use the largest image dimensions as the map for all of the subsequent images. We do this because the 3D plot will not work unless we use the same size map. We then offset the images by one for each image. Now, one can scale the distance between image layers using the \emph{daspect()} function of Matlab to scale the Z (Energy) in order to resolve the layers. 
\begin{lstlisting}
%# plot each slice as a texture-mapped surface (stacked along the Z-dimension)
for k=1:size(im)
      surface('XData',X-0.5, 'YData',Y, 'ZData', Z.*-k, ...
        'CData',im{k}, 'CDataMapping','direct', ...
        'EdgeColor','none', 'FaceColor','texturemap')
end
\end{lstlisting}
\section{Analysis tools of \textsc{SpectroLab}}
\begin{figure*}[!htbp]
 \centering
    \includegraphics[width=0.9\linewidth]{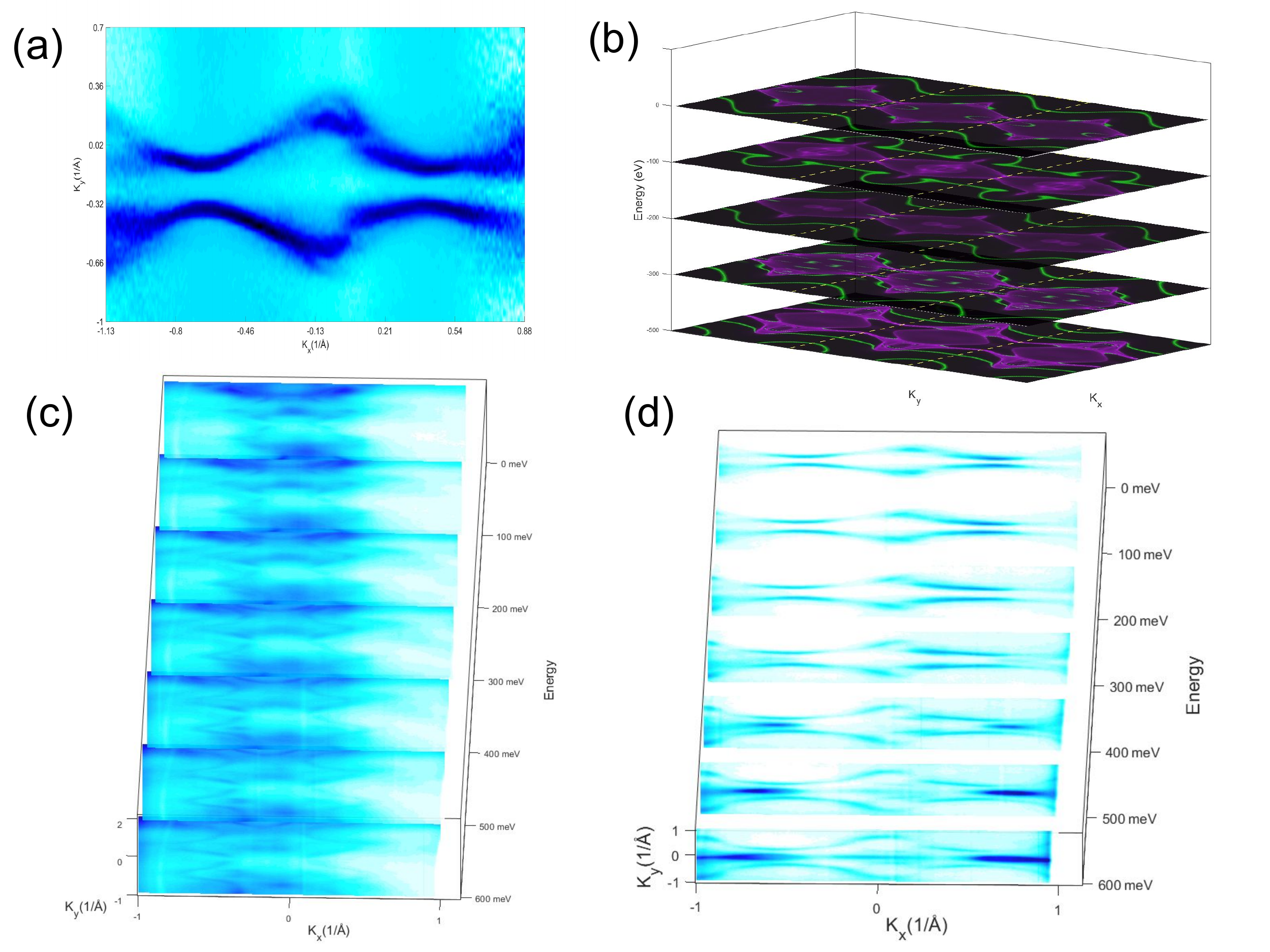}
\caption{\textbf{Example of functions} (a) Result of the fermisurface() function. (b) Result of the stacked\_theory() program. (c) Output of the stackedECP() function in the Hfp$_2$ system. (d) Output of the stackedECP() function in another dataset in the HfP$_2$ system.}
\label{multi}
\end{figure*}

Here we included an overview of the uses of \textsc{SpectroLab}. The fermi level can be found with the \emph{fermisurface()} [Fig. \ref{multi}(a)] and the \emph{ypos} variable
\begin{table}[ht]
\caption{Main tools of \textsc{SpectroLab}}
\begin{tabular}{l|l}
\hline
Program             & Description\\ \hline \hline
Kx Cuts             & Creates $k_x$ momentum slices \\ \hline
Ky Cuts             & Creates $k_y$ momentum slices \\ \hline
Plot\_KX\_KY        & For quick check of data analysis \\ \hline
Fermi Surface       & Draws Fermi surface from 3D map \\ \hline
Momentum 2D         & 2D ARPES cuts \\ \hline
Stacked plot        & Slices and stacks 3D momentum cuts  \\ \hline
Stacked theory      & Stack theory similar to stacked plot \\ \hline
Density Curves      & EDC and MDC \\ \hline
Energy Contour plot & ECP in exact steps \\ \hline
Volumetric plot     & 3D volumetric plot \\ \hline
Dispersion          & photon energy dependence  \\ \hline
Inner core          & Analysis of inner core spectra \\ \hline \hline
\end{tabular}
\label{Tools}
\end{table}
\subsection{Graphical User Interface (GUI)}
\begin{figure*}[!htbp]
 \centering
    \includegraphics[width=0.9\linewidth]{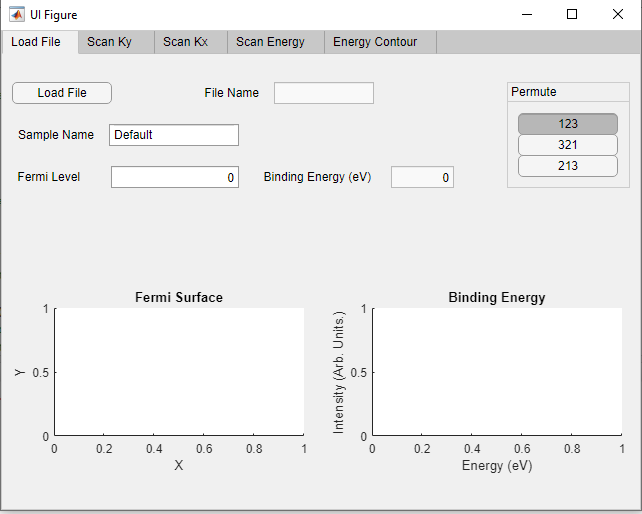}
\caption{\textbf{Graphical User interface} (a) The main page the GUI figure with the main release of \textsc{SpectroLab}.}
\label{GUI}
\end{figure*}

We include a graphical user interface (GUI) [Fig. \ref{GUI}] that can be used as a general tool, more advance data selection and analysis will need to be done in code.
\subsection{Constant Energy Contours}
We include a function to extract constant energy contours (ECs) from a 3D band mapping. StackedECP() [Fig. \ref{multi}(c,d) is a feature that calculates ECs and stacks them on top of each other in order to visualize the bands dispersing in 3D space. The energy contour functions do the same, however the ECs are presented in a 2D format.
\subsection{Volumetric plot}
In the volumetric plot function, a 3D band mapping is interpreted with an alpha map (transparency map). This allows for all bands to be mapped in a 3D space. In addition, users can rotate and zoom in to further examine the electronic structure.
\subsection{Momentum cuts}
\begin{figure*}[!htbp]
 \centering
    \includegraphics[width=0.9\linewidth]{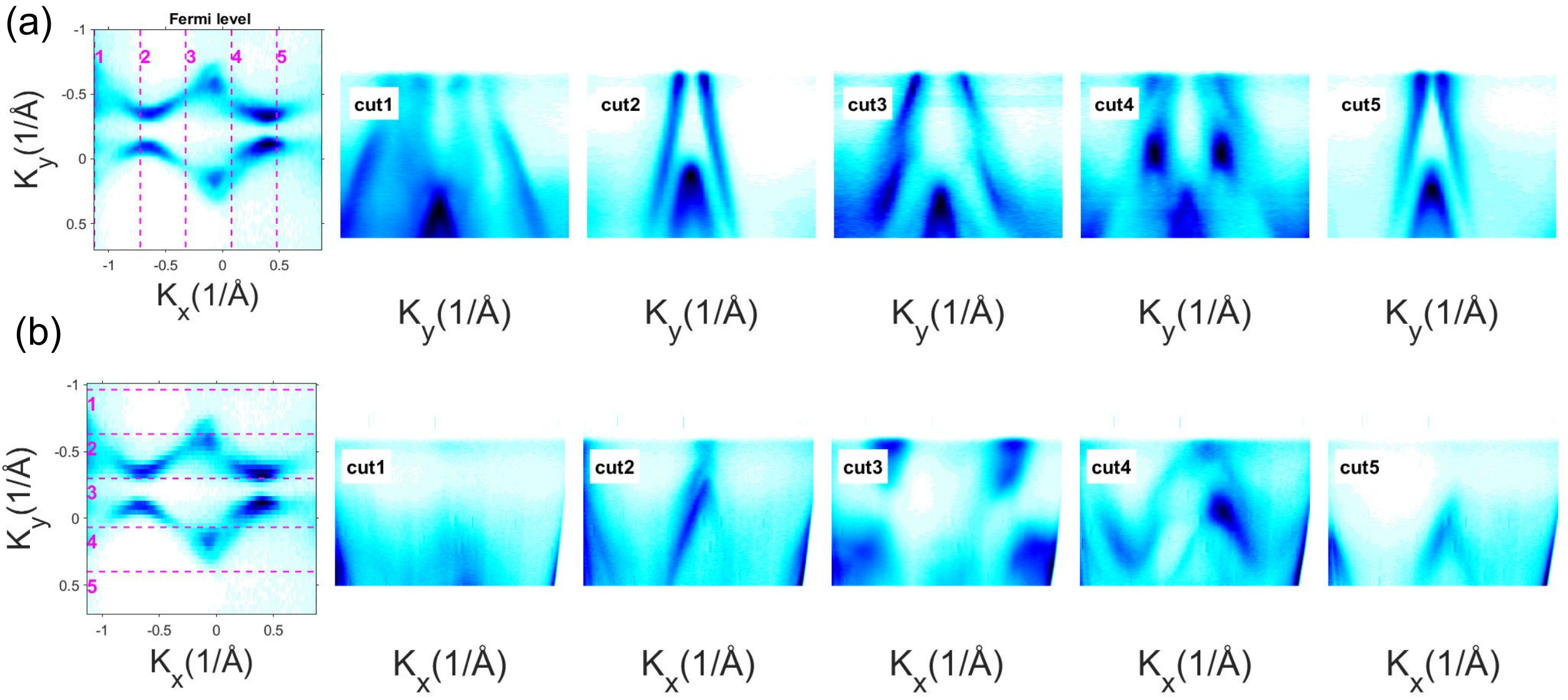}
\caption{\textbf{$k_x $ and $k_y$ momentum cuts} (a) Fermi surface (rightmost panel) and $k_x $ momentum cuts (cut1-cut5) taken from a 3D mapping by using the kxcuts function.
 (b) same as (a) but for $k_y$ momentum cuts}
\label{kxky}
\end{figure*}

We have included several methods to aid the taking $k_x$ and $k_y$ momentum cuts from a 3D band mapping. Plot\_KX\_KY, Ky cuts (Fig. \ref{kxky}(a)), and Kx cuts  (Fig. \ref{kxky}(b)) are all features that allow for one to construct these plots. Plot\_KX\_KY is a debug tool but can still be used to accomplish this feature.

\subsection{Energy \& momentum distribution curves}
\begin{figure*}[!htbp]
 \centering
    \includegraphics[width=0.8\linewidth]{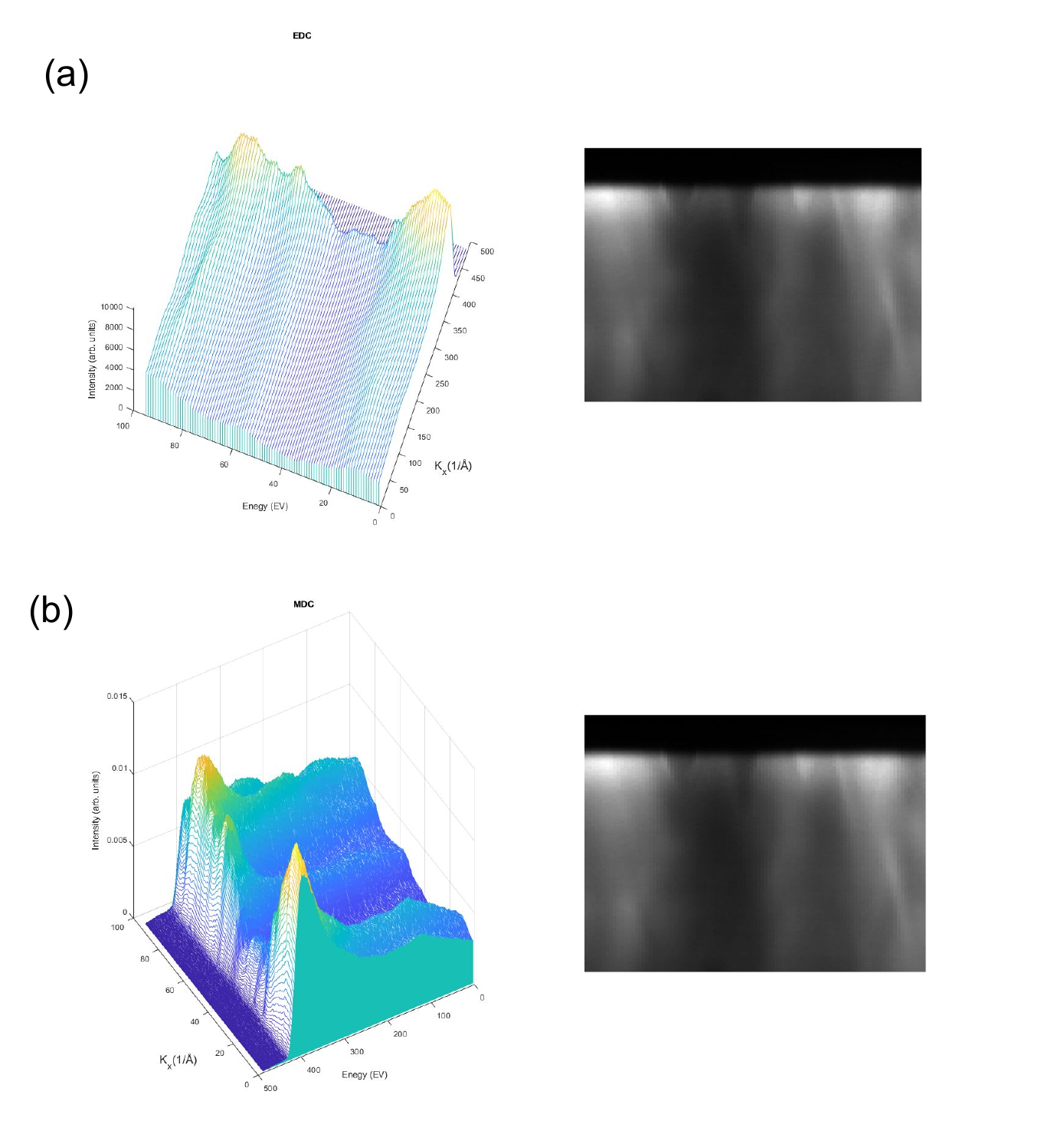}
\caption{\textbf{Energy \& Momentum Distribution Curves} (a) EDC curve (b) MDC curve}
\label{em}
\end{figure*}

Energy distribution curves (EDCs) and momentum distribution curves (MDCs) are important tools for the analysis of ARPES data. Here, we construct programs that can create 3D EDC [Fig. \ref{em}(a)] and MDC [Fig. \ref{em}(b)] plots similar to those found in previous literature. The advantage of these types of plots is that they are able to give a 3D interpretation of the EDC and MDC which will allow for better understanding of the intensity at certain points (such as discerning if there is a band gap or weak feature in the band gap).

\subsection{Photon energy dependence}
\begin{figure*}[!htbp]
 \centering
    \includegraphics[width=0.9\linewidth]{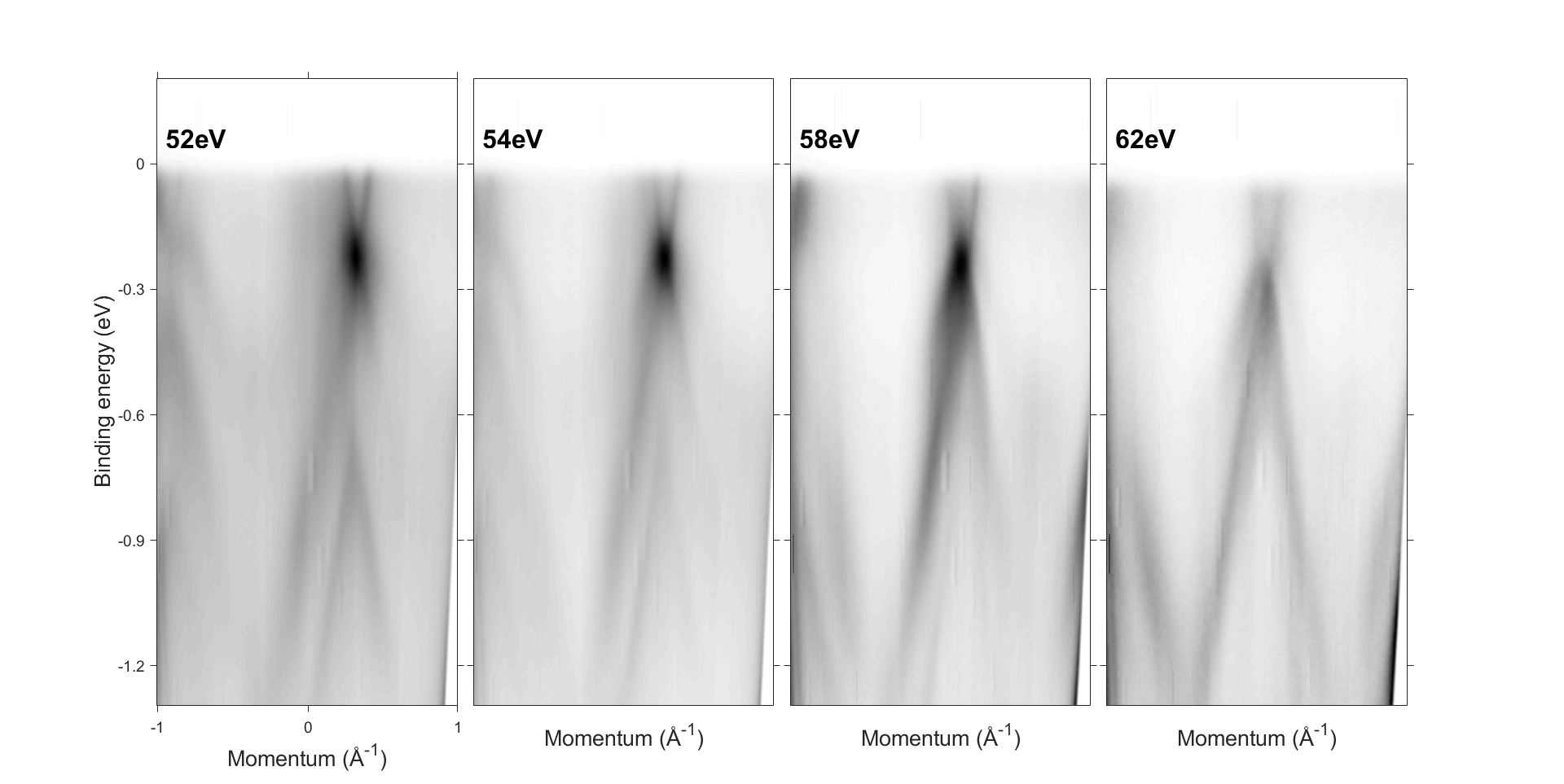}
\caption{\textbf{Dispersion map} A dispersion map of HfP$_2$ resolved with the dispersion() function}
\label{disp}
\end{figure*}

Dispersion allows for one to interpret photon energy dependent cuts (not a mapping of photon energy) to include for interpretation or publication. The Dispersion map function can interpret an arbitrary amount of cuts specified by the user. [Fig. \ref{disp}]

\subsection{Stacked theory}
Stacked theory is a function that allows one to import theoretical DFT energy contours and plot them in a way to compare to the \emph{stackedECP()} function [Fig. \ref{multi}(b)]. There is also a feature that allows for plotting theory in one of the energy contour functions. 
%
\section{Installation and usage}
In this section, we present how to install and use \textsc{SpectroLab}.
\subsection{Get \textsc{SpectroLab}}
\textsc{SpectroLab} is an open source software package distributed under the MIT License. The code can be downloaded directly from the public code repository: \href{https://github.com/christophersims/SpectroLab}{github.com/christophersims/SpectroLab}.
\subsection{Installation and running the code}
The code requires no installation, however, the folders that contain the code must be added to the path in order for them to run correctly (or the user could place all of the .m files in one directory). The user must have a Matlab version year of at least 2018 with the image toolbox add-on installed.

The code can be run by simply specifying the file path of the data one wishes to analyze. If there are no formatting errors the code should run without errors.
\subsection{File formats}
For ARPES there are three major file formats used (from the Igor suite), \textit{.txt}, \textit{.ibw}, and \textit{.pxt}. Important data is lost in the \textit{.txt} format and the \textit{.pxt} format is compressed and cannot be decompressed outside of the Igor program. Therefore, this program only uses the Igor Binary Wave format (\textit{.ibw}). In addition, the program only accepts files in the format ($k_x$,Energy,$k_y$), however, the file can be reformatted within the program.\\


\section{Examples}
\subsection{Incorrectly formatted input}
\begin{figure*}[!htbp]
 \centering
    \includegraphics[width=0.9\linewidth]{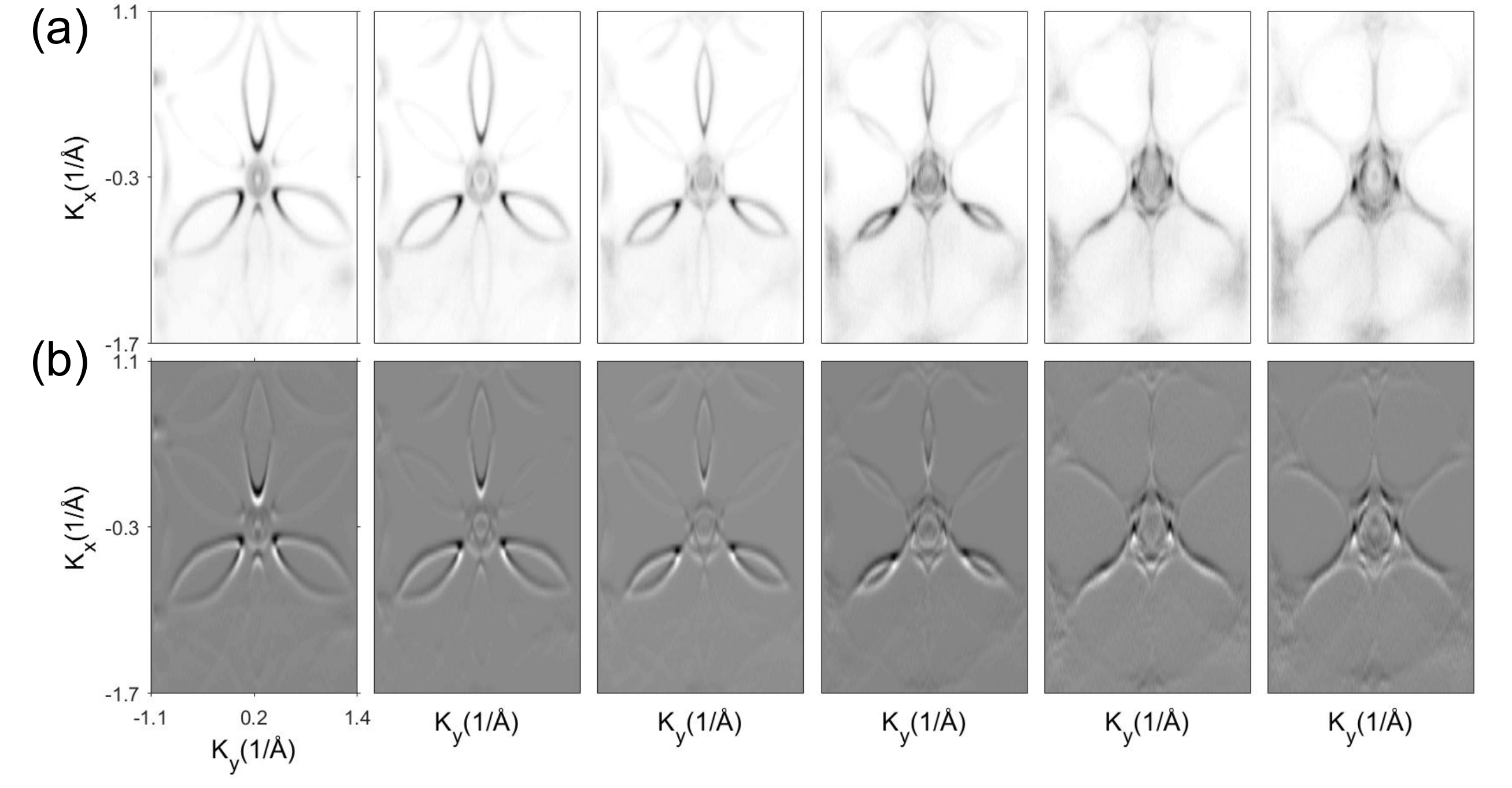}
\caption{\textbf{Hf$_2$Te$_2$P} (a) Normal energy contour (b) 2nd derivative energy contour}
\label{hf}
\end{figure*}

\begin{figure*}[!htbp]
 \centering
    \includegraphics[width=0.9\linewidth]{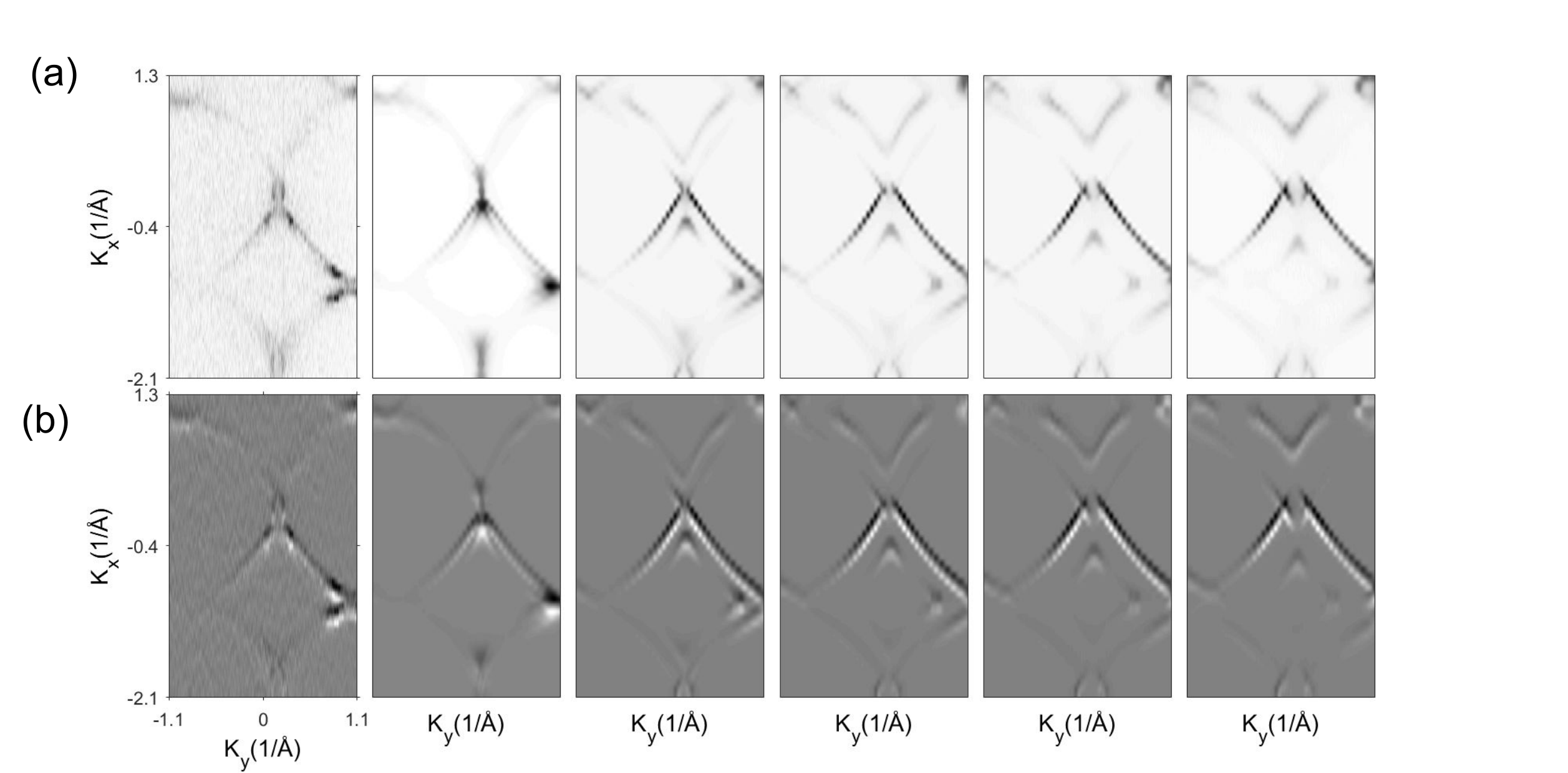}
\caption{\textbf{ZrSiS Energy Contour} (a) Normal energy contour (b) 2nd derivative energy contour}
\label{zr}
\end{figure*}

If the input of the code in not formatted correctly (in the ($k_x$,Energy,$k_y$) format) it is possible to still run the code by utilizing Matlab's permute() function. This function allows for one to swap the matrices in order for the program to run correctly. In addition to using the permute the variables in variables() must also be changed in order to match the code. In figure , we present data that was formatted incorrectly but was able to run in the program and produce good results. Here we present the data of Hf$_2$Te$_2$P (Fig. \ref{hf}) and ZrSiS (Fig. \ref{zr}) which were imported using the permute function.

\subsection{Plotting theory}
Our implementation of plotting theory is currently an image importer, however, we have the code that allows for the input if the surface states and Fermi surface generated with the iterative surface green functions\cite{Sancho1985} imported from Wanniertools\cite{Wu2018}. The calculations are not included in this package and must be calculated with tools such as Quantum Espresso\cite{Giannozzi2009}, VASP\cite{Kresse1993}, Abinit\cite{Gonze2009}, Wannier90\cite{Mostofi2008}, etc. In figure \ref{multi}(b), we show the capability so show stacked theory plots in order to coincide with the stackedECP() function. In addition, we have included a ECP function that also allows for the inclusion of theoretical calculation figures (see supplementary).
\subsection{Volumetric plot}
\begin{figure*}[!htbp]
 \centering
    \includegraphics[width=0.9\linewidth]{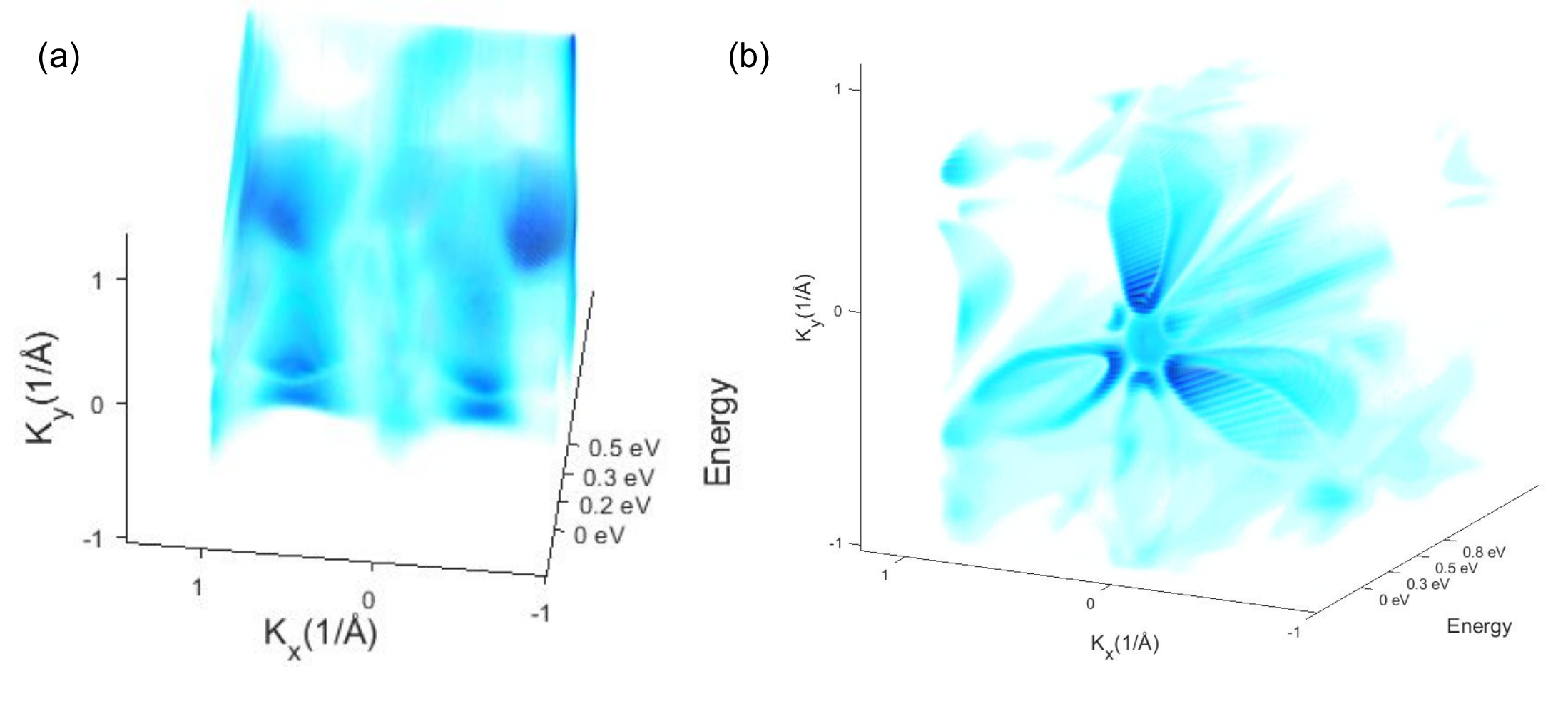}
\caption{\textbf{Volumetric plot} (a) volumetric plot of correctly formatted code. (b) Volumetric plot of incorrectly formatted input. There is no errors in the incorrectly formatted 3D band mapping contour}
\label{vol}
\end{figure*}

Here we present a volumetric plot of both a 3D band mapping correctly formatted Fig. \ref{vol}(a) and an incorrectly formatted (corrected in code) in Fig. \ref{vol}(b). This method is a powerful tool that will enable researchers to visualize bands in a 3D space.

\section{Conclusion}
In conclusion, we have designed a Matlab based program that can be used to interpret spectroscopic data. Included in this tool are multiple functions which can be used for the analysis of ARPES data. To demonstrate the robustness of this package we show ARPES analysis of several data sets. This package can be used as a standalone with no other dependency on other programs. The features of this analysis suite is extensive with the addition of new analysis techniques such as the 3D volumetric plot. In the future, we aim to integrate the analysis of XRD and STM data into the \textsc{SpectroLab} suite.

\section*{Supplemetary Material}
See Supplementary material for further figures of \textsc{spectrolab}'s output in the HfP$_2$, Hf$_2$Te$_2$P, and ZrSiS systems.
\section*{Acknowledgments}
The authors would like to thank the Matlab community which provided several methods and programs used in \textsc{SpectroLab}.

C.S. acknowledges the generous support of the Purdue ECE ASIRE fellowship and the National GEM consortium fellowship\\

Correspondence and requests for additional functionality should be addressed to C.S. \\(Email: Sims58@purdue.edu).
%

\newpage



\end{document}